\documentclass[12pt]{article}

\newcommand{\be}[1]{\begin{equation}\label{#1}}

\newcommand{\ee}{\end{equation}}

\newcommand{\omg}{\Omega}
\newcommand{\com}{\hbox{\rlap{\hskip2.5pt \vrule height 6pt width0.6pt}C}}

\newcommand{\s}{\sigma}

\title{\bf Finitary Spacetime Sheaves} 

\author{I. Raptis\footnote{Department of Mathematics, 
University of Pretoria, Pretoria 0002, Republic of South Africa; e-mail: 
iraptis@math.up.ac.za}}
  
\date{}

\begin{document}

\maketitle                                                               

\begin{abstract}

The notion of finitary spacetime sheaves is introduced based on locally 
finite approximations of the continuous topology of a bounded region of a 
spacetime manifold. Finitary spacetime sheaves are seen to be  
sound mathematical models of approximations of continuous spacetime 
observables.

\end{abstract}

\section{\normalsize\bf Introduction cum Physical Motivation}

\sloppy{The present paper associates with a finitary substitute $F_{n}$ of a bounded 
region $X$ of a continuous spacetime manifold $M$ (Sorkin, 1991), a 
collection $S_{n}$ of appropriately defined continuous functions on $X$, which, 
as a space on its own, is seen to be (locally) homeomorphic to $F_{n}$, thus 
technically 
speaking, a sheaf over $F_{n}$ (Bredon, 1967). This finitary spacetime sheaf 
is denoted by $S_{n}(F_{n})$.  
  
Then, we consider an inverse system ${\cal{K}}=<F_{n}({\cal{U}}_{n})>$ of finitary  spacetime 
substitutes, derived from a net ${\cal{L}}=<{\cal{U}}_{n}>$ of locally 
finite open covers of $X$ as in (Sorkin, 1991), and the corresponding 
inverse system ${\cal{N}}=<S_{n}(F_{n})>$ of finitary  
spacetime sheaves associated with each element of ${\cal{K}}$. We 
show that as the 
elements $F_{n}$ of ${\cal{K}}$  get more refined (in a sense to be defined), 
their 
corresponding sheaves $S_{n}$ in ${\cal{N}}$ `converge' to $S(X)$-the sheaf 
of (germs of) continuous functions on $X$.

The central physical idea we wish to model by the finitary spacetime sheaves  
$S_{n}$ is `locally finite approximations of the continuous spacetime 
observables on $X$'.
In more detail, we intuit that as the continuous topology of $X$ can be  
finitely (or coarsely) approximated by the finitary topologies $F_{n}$, 
so can the continuous maps on it, that constitute the sheaf $S(X)$, be 
effectively approximated by the finitary sheaves $S_{n}$. Since only the 
continuous ({\it ie}, $C^{0}$) topological structure of $X$ concerns us here, 
that is to say, we consider 
the continuous topology of spacetime to be its sole physically significant 
property, 
the physical spacetime observables that are 
of relevance are precisely the continuous functions on it. Then, the maps 
in each finitary sheaf $S_{n}$ over the locally finite substitute $F_{n}$ of  
$X$ represent sound coarse approximations of the continuous 
observables in $S(X)$ in the sense that, `in the limit of infinite          
resolution' of the $F_{n}$s into $X$, they effectively 
reproduce them. The latter may be formally written as 
$\lim_{n\rightarrow\infty} S_{n}(F_{n})\equiv
S_{\infty}(F_{\infty})\simeq S(X)$\footnote{`$\simeq$' denoting 
`homeomorphic to'.} ($S_{n}(F_{n})\in{\cal{N}}$), with an explanation of 
this limiting procedure pending.

The present paper is organized as follows: in section 2, a brief 
reminder of finitary substitutes of a bounded region $X$ of a continuous 
spacetime manifold $M$ is given. 
We also allude, without proof, to the `inverse limit' topological 
space $F_{\infty}$ to which an appropriately defined `inverse system'   
${\cal{K}}$ of such substitutes derived from a net 
${\cal{L}}=<{\cal{U}}_{n}>$ 
of locally finite open covers of $X$, converges `at 
maximum resolution of $X$ by ${\cal{U}}_{\infty}$'. 
This topological space is essentially homeomorphic to $X$. For detailed 
proofs the reader is referred to (Sorkin, 1991).

In section 3, a space $S$ that is locally homeomorphic to $X$ is defined. 
This is the sheaf $S(X)$ of continuous functions on $X$.

In section 4, finitary spacetime sheaves $S_{n}$ of continuous functions 
on $X$ associated with the finitary substitutes $F_{n}$ of section 2, 
are defined constructively. 

Like their finitary `domains' $F_{n}$, the finitary spacetime sheaves 
$S_{n}$ also form an inverse system ${\cal{N}}$ having an inverse limit 
topological space $S_{\infty}(F_{\infty})$ that is essentially homeomorphic 
to the sheaf $S(X)$ of section 3. This limiting procedure is briefly 
presented in section 5 
by simulating in ${\cal{N}}$ Sorkin's proof that the inverse system  
${\cal{K}}$ `converges' to $X$. This is evidence of 
the soundness of the finitary sheaves as models of locally 
finite approximations of the continuous functions on $X$-the observables  
of the continuous spacetime topology of $X$.

In the concluding section 6 we discuss the physical significance of the 
elements of finitary spacetime sheaves as the locally finite approximations 
of the continuous observables on the region $X$ of the spacetime manifold 
$M$. In view of a recent definition of quantum causal sets (Raptis, 2000), 
partially motivated by the poset finitary substitutes of continuous spacetime 
topologies in (Sorkin, 1991), their quantum algebraic analogues in (Raptis and 
Zapatrin, 2000) and their causal relatives in (Bombelli {\it et al.}, 1987), 
we entertain the idea of a finitary spacetime sheaf 
of quantum causal sets, as well as the possibility that this structure be  
curved, thus serve as a sound model of some sort of `finitary quantum 
gravity'. However, the analytic development of this possible application of 
finitary spacetime sheaves will have to be postponed for another paper 
(Mallios and Raptis, 2000).}  

\section{\normalsize\bf Finitary Spacetime Substitutes Revisited}

\sloppy{In (Sorkin, 1991), with finite open covers ${\cal{U}}_{n}$ of a bounded 
region $X$ of a continuous spacetime manifold $M$\footnote{A subset of a 
topological space is said to be bounded if its closure is compact. Our 
$X$ in $M$ is bounded in this sense.}, finite topological 
spaces $F_{n}$ were associated by the following `equivalence algorithm': 
two points $x$ and $y$ in $X$ are said to be equivalent with respect to 
the open cover ${\cal{U}}_{n}$ (write $x\stackrel{{\cal{U}}_{n}}{\sim} y$),  
when $\forall U\in{\cal{U}}_{n}:\, x\in U\Leftrightarrow y\in U$.

Denoting by $\Lambda(X)$ the smallest open neighborhood in the 
subtopology ${\cal{T}}_{n}$ of $X$ generated by the open sets $U$ in 
${\cal{U}}_{n}$\footnote{That is to say, ${\cal{T}}_{n}$ consists of 
arbitrary unions of finite intersections of the open sets $U$ in 
${\cal{U}}_{n}$.}

$$\Lambda(x):=\bigcap\{ U\in{\cal{U}}_{n}|~x\in U\} ,$$ 

\noindent one may alternatively define $x\stackrel{{\cal{U}}_{n}}{\sim} y$ as

$$(x\rightarrow y)\wedge(y\rightarrow x),$$

\noindent where $x\rightarrow y$ stands for $x\in\Lambda(y)$. 

The quotient space 
$X/\stackrel{{\cal{U}}_{n}}{\sim}=:F({\cal{U}}_{n})\equiv F_{n}=\{ [x]\}$ 
consisting of equivalence classes $[x]$  
of points $x$ in $X$ relative to its finite open covering ${\cal{U}}_{n}$, 
is  seen to be a $T_{0}$ topological space having the structure of a poset 
(Sorkin, 1991). With this  
$T_{0}$-quotienting of $X$ by the equivalence relation 
$\stackrel{{\cal{U}}_{n}}{\sim}$ to $F_{n}$, that is 
to say, the substitution of $X$ by its finitary approximation $F_{n}$ relative 
to ${\cal{U}}_{n}$, a continuous function  
$f_{n}$ from the subtopology ${\cal{T}}_{n}$ of $X$ to $F_{n}$, may be  
associated (Sorkin, 1991). $f_{n}$ is continuous in the usual sense that 
open sets in 
$F_{n}$\footnote{Defined with respect to the partial order relation 
$\rightarrow$ in $F_{n}$ as the sets that can be obtained as unions 
of the following basic open sets: $\forall x\in F_{n}:~\Lambda(x):=\{ y\in 
F_{n}|~ y\rightarrow x\}$; where we have used the points $x$ instead of the 
equivalence classes $[x]$ where they belong for simplicity of notation.} are 
mapped by $f_{n}^{-1}$ to open subsets of $X$ in ${\cal{T}}_{n}$. 

In (Sorkin, 1991), a net ${\cal{L}}$ of open covers $\{{\cal{U}}_{n}\}$ is 
also considered. For 
every pair ${\cal{U}}_{i}$ and ${\cal{U}}_{j}$ of locally finite open covers 
of $X$ in ${\cal{L}}$, there is a `finer' open cover 
${\cal{U}}_{k}\in{\cal{L}}$ such that ${\cal{U}}_{i},{\cal{U}}_{j}\subset
{\cal{T}}_{k}$. With the net ${\cal{L}}$ and the $f_{n}$s corresponding 
its elements to the $F_{n}$s, the inverse system ${\cal{K}}=<F_{n}>$ of finitary 
substitutes of $X$ may be derived. Here too, $F_{i}\preceq F_{k}$ may be 
taken to stand for $F_{i}\subset {\cal{T}}_{k}$, and means that the open sets 
in the poset $F_{i}$ are contained in the finitary poset topology 
${\cal{T}}_{k}$ generated by the open sets in the poset $F_{k}$.   

For the pair $i$ and $k$ of indices above, the `finer' 
relation may be denoted as $i\preceq k$ and it is 
seen to be a partial order between the locally finite open  
covers of $X$ in ${\cal{L}}$, or their associated finitary substitutes in 
${\cal{K}}$. Intuitively, the procedure of `refinining' an open 
cover ${\cal{U}}_{i}$ to a finer ${\cal{U}}_{j}$ ($i\preceq j$), corresponds 
to adding more and finer or `smaller' open subsets of $X$ to the open cover     
${\cal{U}}_{i}$ to obtain ${\cal{U}}_{j}$, thus, in a way, it represents  
the employment of higher power of resolution in operations of 
determination of the topological structure of $X$\footnote{We will return to 
this issue at the end of section 5 and in section 6.}. 

The net ${\cal{L}}$ 
of finite open covers of $X$ provides the basis for the definition of 
${\cal{K}}$ as an inverse-system (Sorkin, 1991). For every pair 
${\cal{U}}_{i}$ and ${\cal{U}}_{j}$ in ${\cal{L}}$ (or the corresponding pair 
of posets in ${\cal{K}}$), with ${\cal{U}}_{i}\preceq{\cal{U}}_{j}$ (or 
$F_{i}\preceq F_{j}$), one defines a map $f_{ij}:~F_{j}\rightarrow F_{i}$, 
which is seen to be a unique continuous surjection. 
  
The central result from (Sorkin, 1991) is that the inverse system 
${\cal{K}}$ 
`converges' to $F_{\infty}$ at the (inverse) limit of maximum resolution 
({\it ie}, formally, as $n\rightarrow\infty$) of the finite open covers of 
$X$ in ${\cal{L}}$. In particular, $F_{\infty}$ is essentially $f_{\infty}$-homeomorphic 
to $X$, so that, effectively, the maximally refined finitary substitute of 
$X$ is topologically indistinguishable from (or equivalent to) it. 
This description of 
$f_{\infty}$ as being `essentially a homeomorphism' between $X$ and 
$F_{\infty}$,   
pertains to the fact, shown in (Sorkin, 1991), that $X$ is 
$f_{\infty}$-embedded in $F_{\infty}$ as a dense subset. Then, Sorkin 
shows how to `discard' from $F_{\infty}$ the collection of its `extra-points' 
$y$ that are `infinitely close' to those of $f_{\infty}(X)$, thus establish  
$f_{\infty}$ as a homeomorphism between $\hat{F}_{\infty}=F_{\infty}/\{ 
y\}$ and $X$\footnote{For the purposes of the present paper, we need not 
explain Sorkin's proof of the inverse limit in more detail. A brief 
outline of the basic steps of the proof is given in section 5 where we 
argue that $S_{n}(F_{n})$ in ${\cal{N}}$ converge to $S(X)$.}.

The last result from (Sorkin, 1991) of interest to us here is that the 
tiling 
of $F_{\infty}$ by the open sets in a particular covering ${\cal{U}}_{i}$ 
of $X$, symbolized as $f_{i\infty}^{-1}(x)$ ($\forall x\in 
F_{i}$)\footnote{Where $f_{i\infty}:~ F_{\infty}\rightarrow F_{i}$ can be 
thought of as the continuous finite approximation of 
$F_{\infty}$ by $F_{i}$.}, 
becomes arbitrarily fine as $i\rightarrow\infty$. In other words, the 
open sets get `smaller' and `smaller' as the `experimenter' employs higher 
power to resolve $X$ into its points ({\it ie}, ideally, to determine 
or localize individual spacetime events in $X$). 

In the next section we present a space $S$ that, similarly to $F_{\infty}$, 
is (locally) homeomorphic to $X$.}  
   
\section{\normalsize\bf ${\bf S(X)}$-The Sheaf of Continuous Functions over 
${\bf X}$}

\sloppy{We want to organize the observables on the bounded spacetime region $X$ 
of the continuous manifold $M$ into a space $S$ that, for all practical 
purposes, is topologically indistinguishable from ({\it ie}, homeomorphic to)  
$X$. If we succeed in this, then the information about the topology of 
spacetime encoded in the points of $X$, which serve as the `carriers of 
its topology'\footnote{See section 6 for the discussion of this physical 
role of the points of $X$.} (Sorkin, 1991), will be the same as that of our 
observations of them, at least locally\footnote{That is to say, `about 
every point-event $x$ in $X$'.}, and our description of the topological 
relations between spacetime events will be the same as that of the topological 
relations between our observations of them. Thus, the problem of localization 
of spacetime events will be effectively translated to the more operationally 
sound problem of localization of our observations of them\footnote{Again, see 
the concluding section 6 for more discussion on this.}. 

Now, there is such an organization of the continuous 
functions on $X$ called a sheaf. 
Below, we first introduce the notion of a 
presheaf $P(X)$ of functions defined on the open sets of $X$, 
then we endow this 
presheaf with a topology that is locally equivalent ({\it ie}, homeomorphic) 
to that of $X$. This process converts the presheaf $P(X)$ into the sheaf $S(X)$ of (germs 
of) continuous functions on $X$ and is called `sheafification'. Only 
the basic definitions from sheaf theory, that will help us define 
finitary spacetime sheaves in the next section, are given below. For a 
more detailed treatment of sheaves, the reader is referred to (Bredon, 1967).    

A presheaf $P$ on our region $X$ of the continuous spacetime manifold $M$ 
is an assignment to each open subset $U$ of $X$ of a set $P(U)$ and to each 
pair $U$, $V$ of open subsets of $X$ of a `restriction map' $\rho_{U,V}:~ 
P(V)\rightarrow P(U)$, so that: $\rho_{U,U}=id$, and 
$\rho_{U,V}\rho_{V,W}=\rho_{U,W}$, with $id$ the identity map and 
$U\subset V\subset W\subset X$. 
One may think of the presheaf $P(X)$ as collections of functions defined on open subsets 
of $X$ which $\rho$-reduce to one another when their respective domains of 
definition are nested by inclusion\footnote{This nesting by inclusion of the 
open subsets of $X$ is a partial order on the collection of all open subsets 
of $X$.}.

To sheafify or `topologize' the presheaf $P(X)$ to the sheaf $S(X)$, we 
embed each $P(U)$ in the presheaf to the collection $\Gamma(U,S)$ of  
sections of continuous functions on $U\subset X$ by the map 
$\s_{U}:~ P(U)\rightarrow 
\Gamma(U,S)$, that is to say, in some sense we select from $P(U)$ the continuous 
maps on $U$. $\s_{U}$ commutes with the $\rho$-restrictions of open sets.
Then, for every $x$ in $X$ we define the equivalence 
relation $\stackrel{x}{\sim}$ between the elements of the sets $P(U)$ and 
$P(V)$ ($U\cap V\not=\emptyset$) as follows

$$f\stackrel{x}{\sim}g,~(f\in P(U),g\in P(V))\Leftrightarrow 
\rho_{W,U\cap V}(f)=\rho_{W,U\cap V}(g),~ (x\in W\subset U\cap V).$$

\noindent With the definition of $\stackrel{x}{\sim}$, one may define the
stalk of the sheaf $S$ over $x$ as the following equivalence class 
of continuous functions at $x$

$$S_{x}:=\bigcup\{ P(U)|~ x\in U\}/\stackrel{x}{\sim}~\equiv \vec{\lim}\{ 
P(U)|~ x\in U\}\, , $$

\noindent with $\vec{\lim}$ denoting `direct limit'\footnote{This limit effectively 
yields the `smallest' class of functions in the presheaf $P$ over the `finest' 
neighborhood of $x\in X$ (that is also included in 
both $U$ and $V$). This direct limit effectively corresponds to maximum 
localization/resolution of $S(X)$ into its stalks $S_{x}$ over the points $x$ 
of $X$, and it is dual to the inverse limit procedure employed to 
resolve $X$ into its points $x$ in (Sorkin, 1991)-read the physical 
significance of finitary spacetime sheaves in section 6.}. As a                                       
non-topologized set, the sheaf $S(X)$ may be expressed as a disjoint union 
(or direct sum) of its stalks $S(X)=\bigcup_{x\in X}S_{X}$\footnote{Thus, 
in some sense, the stalks $S_{x}$ of $S(X)$ are its `points', like the $x$ in 
$X$-see last footnote and read the physical significance of finitary spacetime 
sheaves in section 6.}.

We may now endow $S$ with the following topology ${\cal{T}}$: let $f$ be a 
member of the presheaf $P(U)$ and $x$ a point in $U$, then the germ of $f$ at 
$x$, $[f]_{x}$, is the $\stackrel{x}{\sim}$-equivalence class of $f$. 
A basis for the topology of $S$ consists of open sets of the following sort: 
$(x,[f]_{x})$ ($x\in U$).

Now, $\Gamma(U,S)$ above is the set of continuous sections of the sheaf $S(X)$ over 
its open subset $U$, that is to say, the set of continuous maps $s:~ 
U\rightarrow S$ such that, locally ({\it ie}, point-wise in $X$), they map 
each point $x$ in $X$ to (an element $[f]_{x}$ of) the stalk $S_{x}$ over it. Also, the selection 
$\s_{U}$ in  $\Gamma(U,S)$ of the map $f$ in $P(U)$ above reduces locally to 
the germ of $f$ at $x$, that is: $\s_{U}(f)(x)=[f]_{x}\in S_{x}$ $(x\in U)$. It 
follows that, as a topological space ${\cal{T}}(S)$, the sheaf $S(X)$ is 
generated by the (germs of) sections of the continuous maps on $X$, so that 
one can easily show that the `projection map' $\pi:~ S(X)\rightarrow X$, given 
locally by $\pi(x,[f]_{x})=x$, is a local homeomorphism; hence, the slogan 
that `a sheaf is a local homeomorphism'. Equivalently, one can verify that 
every (germ of a) section of a continuous map in $\Gamma(U,S)$ is also such a 
local homeomorphism of $X$ to $S(X)$\footnote{Hence the equivalent 
slogan that `a sheaf is its sections'.}, so that for such a section 
$s\in\Gamma(U,S)$, the composition of $s$ with $\pi$ corresponds to 
the identity map $id:~ S(X)|_{U}\rightarrow S(X)|_{U}$ ({\it ie}, $s\circ\pi\equiv 
id_{S(X)|_{U}}\Leftrightarrow s\equiv\pi^{-1}$), where $S(X)|_{U}$ is the restriction of the sheaf $S(X)$ to 
the open subset $U$ of $X$\footnote{We may add here that usually the stalks $S_{x}\in S(X)$ 
are assumed to have some kind of algebraic structure, so that the algebraic 
operations that define it respect the `horizontal continuity' of the base 
space $X$ ({\it 
ie}, they satisfy `compatibility conditions' with the topology of the 
underlying space $X$). It must be noted that the continuous functions on 
$X$ form indeed an algebra, the prototype being $C^{0}(X,\com)$-the algebra 
of complex valued continuous functions on $X$. It is tacitly assumed that 
the operations in $C^{0}(X,\com)$ respect the base topology in a sheaf of 
such algebras over $X$. In the present paper we are not interested in  
the algebraic structure of the stalks of the (finitary) sheaves. In section 
6 however, we mention finitary spacetime sheaves whose stalks have some 
specific algebraic structure of special importance to the physical situation 
that they are employed to model.}.

In the next section we define finitary spacetime sheaves $S_{n}(F_{n})$ as
the finitary substitutes of the continuous sheaf $S(X)$ in a way analogous 
to how the $F_{n}$s were seen to be the locally finite poset 
substitutes of the continuous topological space $X$ in (Sorkin, 1991) and 
section 2.} 
                                                                             
\section{\normalsize\bf Constructive Definition of ${\bf S_{n}(F_{n})}$}

\sloppy{A point in $X$ ({\it ie}, a spacetime event) corresponds to an ideal determination 
of location in spacetime-an ideal measurement of the locus of an event. 
A more pragmatic and operationally sound model of spacetime measurements, one  
taking into account their actual `roughness' or `approximate character', or 
even the `fuzziness' due to the uncontrollable perturbations that  
such realistic acts of measurement inflict on spacetime, is that they, 
at least, determine open sets in $X$ (Sorkin, 1995, Raptis and Zapatrin, 
2000). Since points in $X$ serve as the `carriers of the spacetime topology' 
(Sorkin, 1991)\footnote{See section 6.}, the aforementioned pragmatic determinations of events by 
open sets of $X$ effectively correspond to `approximations of its continuous 
topology'. Thus, the latter may be thought of as the ideal limit-topology an 
experimenter determines at his (also ideal !) maximum power of resolution 
of spacetime into its finest/`smallest' open sets containing its points. 
This was essentially the moral of section 2.

Now our main physical motivation for defining finitary spacetime sheaves 
$S_{n}$ for every finitary approximation $F_{n}$ of the continuous topology 
of $X$, is to similarly approximate the sheaf $S$ of continuous functions 
on $X$, and effectively recover it at the limit of maximum resolution into 
its `ultra-local elements' that in the previous section were seen to be 
its stalks $S_{x}$ over the points $x$ in its base space $X$. 
As we saw in the last section, the stalks of $S$ over the points of 
$X$ consist of the germs of continuous functions on $X$. Thus,  
the main idea is to define our finitary sheaves $S_{n}({F_{n}})$ in 
such a way that their stalks consist of `gross observables' that 
are continuous over the `rough' open subsets of a finite cover 
${\cal{U}}_{n}$ of $X$ and that, in the limit of maximum resolution of the 
${\cal{U}}_{n}$s to $F_{\infty}$ as in section 2, they reduce to the $S_{x}$ 
of $S(X)$. This (inverse) limit will be the subject of the next section.

The finitary spacetime sheaf $S_{n}$,  
associated with the locally finite open cover ${\cal{U}}_{n}$ of $X$, 
is defined as 
follows: take a finitary open cover ${\cal{U}}_{n}$ of $X$. Recall from 
section 2 that  
$\stackrel{{\cal{U}}_{n}}{\sim}$ denotes the following equivalence relation 
between points $x$ and $y$ of $X$ 

$$x\stackrel{{\cal{U}}_{n}}{\sim}y\Leftrightarrow(x\rightarrow \Lambda(y))
\wedge(y\rightarrow\Lambda(x));~\Lambda(X):=\bigcap \{ U\in{\cal{U}}_{n}|~x\in U\}\, ,$$

\noindent according to which, the poset finitary substitute $F_{n}$ of $X$ 
with respect to its locally finite open cover ${\cal{U}}_{n}$, is 
defined by the quotient 
$F_{n}:=X/\stackrel{{\cal{U}}_{n}}{\sim}$ and consists of equivalence
classes $[x]$ of points $x\in X$. Then, force the following 
`collapse' equivalence relation between the stalks $S_{x}=\{[f]_{x}\}$ of 
$S(X)$ over 
the $x$ in $X$, induced by the $\stackrel{{\cal{U}}_{n}}{\sim}$ equivalence 
relation between them, as 

$$x\stackrel{{\cal{U}}_{n}}{\sim}y\Rightarrow  
[f]_{x}\stackrel{{\cal{U}}_{n}}{\equiv}[f]_{y}\, ,$$

\noindent where we have effectively identified the germs of 
continuous functions in the stalks $S_{x}$ and $S_{y}$ of $S(X)$ over the  
$\stackrel{{\cal{U}}_{n}}{\sim}$-equivalent points $x$ and 
$y$ in $X$\footnote{Note that 
all our definitions are implemented `point-wise in $X$'. As noted earlier, 
we hold to the primitive 
intuition that the points of $X$ are `the carriers of its topology'-the basic 
tenet of point set topology (read the physical motivation in section 6 for a 
`physical justification' of this primitive intuition).}. 
This $\stackrel{{\cal{U}}_{n}}{\equiv}$-identification (equivalence relation) 
of the stalks of $S(X)$ has the following physical 
interpretation: for point-events in $X$ that are 
$\stackrel{{\cal{U}}_{n}}{\sim}$-indistinguishable 
at the power of resolution employed to analyze it by ${\cal{U}}_{n}$, 
one may use any of the germs of the continuous observables from $S(X)$, 
residing in the stalks over them, in order to describe their 
`local topological relations'\footnote{Again, read section 6 for more on 
this.}.  

Then, like the $F_{n}$s, we define the                                       
corresponding finitary sheaf  
$S_{n}(F_{n})$, as the following equivalence class of stalks in $S(X)$

$$S_{n}(F_{n}):=S(X)/\stackrel{{\cal{U}}_{n}}{\equiv}\,
=\!\bigcup_{[x]\in F_{n}}\! S_{[x]}.$$                     

It is plain to see that the finitary sheaves $S_{n}(F_{n})$ inherit the 
poset $T_{0}$-topology 
of their corresponding finitary substitutes $F_{n}$ as follows

$$[x]\rightarrow [y]\in F_{n}\Rightarrow [f]_{[x]}\rightarrow [f]_{[y]}\in 
S_{n}(F_{n})\, ,$$

\noindent with the partial order in $S_{n}(F_{n})$ pending a bit of further 
explanation. This explanation can be drawn straightforwardly by considering 
the following `commutative diagram'

$$\begin{array}{rcl}
&{X}\,{\buildrel{f_{n}}\over{\longrightarrow}}\,\,{F_{n}}\cr
&\pi^{-1}\downarrow s \hskip 0.15in s_{n}\downarrow \pi_{n}^{-1}\, .\cr
&{S}\,{\buildrel{{\hat{f}}_{n}}\over{\longrightarrow}}\,\,{S_{n}}\cr
\end{array}$$

\noindent where $s_{n}\equiv\pi_{n}^{-1}$ is the `local homeomorphism' 
from $F_{n}$ to $S_{n}$ that we are searching for. The diagram shows that, 
as a map: $s_{n}=\hat{f}_{n}\circ\pi^{-1}\circ f^{-1}_{n}$, where 
$f^{-1}_{n}$ is the inverse of the bijective correspondence between the 
smallest open subsets $\{\Lambda(x)\}$ containing the points of $X$ with 
respect to its 
locally finite open covering ${\cal{U}}_{n}$ and the 
$\stackrel{{\cal{U}}_{n}}{\sim}$-equivalence classes of points of $X$ 
in $F_{n}$\footnote{Defined in section 2 as the map 
$f_{n}:~{\cal{T}}_{n}\rightarrow F_{n}$. $f_{n}$ as a map 
from $X$ to $F_{n}$ is a continuous surjection, but as a map from 
${\cal{T}}_{n}$ to $F_{n}$ it is a homeomorphism ({\it ie}, a bijective 
continuous map whose inverse is also continuous).}, $\pi^{-1}\equiv s$ is 
the local homeomorphism from $X$ to $S$ as defined in the previous section 
which, in terms of the corresponding smallest open neighborhoods 
$\Lambda(x)$ of points $x$ in 
$X$ with respect to ${\cal{U}}_{n}$, reads: $s:~\Lambda(x)\rightarrow         
\Gamma(\Lambda(x),S)$, and ${\hat{f}}_{n}$ is the one-to-one map defined by 
the equivalence relation $\stackrel{{\cal{U}}_{n}}{\equiv}$ between 
the stalks of $S$ with respect to ${\cal{U}}_{n}$. Thus, as the partial 
order $x\rightarrow y$ (or $[x]\rightarrow [y]$) in $F_{n}$ stands 
(by definition) for 
$x\in\Lambda(y)$\footnote{And it literally stands for the convergence of the 
constant 
sequence $x$ to $y$ (Sorkin, 1991).}, $[f]_{[x]}\rightarrow [f]_{[y]}$ in $S_{n}$ 
is simply interpreted as the following set-theoretic inclusion in $S$: 
$[f]_{[x]}\in\{[f]_{\Lambda([y])}\}=\pi^{-1}(\Lambda(y))=S_{\Lambda(y)}=
\bigcup_{x\in\Lambda(y)}S_{x}$-the restriction of $S(X)$ at 
$\Lambda(y)\subset X$. 
All in all, $s_{n}:~ F_{n}\rightarrow 
S_{n}$ is a local homeomorphism, because, by construction, it is one-to-one
and (locally) preserves the $T_{0}$ order-topology of both $F_{n}$ and $S_{n}$.

This completes the definition of the finitary sheaf $S_{n}$ over the locally 
finite poset substitute $F_{n}$ of $X$. In the next section we argue that, 
as the inverse system ${\cal{K}}=<F_{n}>$ converges to $X$, so its 
derivative ${\cal{N}}=<S_{n}(F_{n})>$ converges to $S(X)$.}                   

\section{\normalsize\bf  The Inverse Limit ${\bf S(X)}$ of ${\bf 
<S_{n}(F_{n})>}$}

\sloppy{In this section we present briefly the basic definitions and follow in little  
detail the main steps in Sorkin's proof of the `convergence' of the inverse 
system ${\cal{K}}$ of 
finitary substitutes $F_{n}$ and the maps $f_{ij}$ between them to a 
space $S_{\infty}$ that is essentially homeomorphic to $X$, and apply 
it for the proof of a similar `convergence' of the inverse system ${\cal{N}}$ 
of the finitary sheaves of the previous section and certain continuous surjections 
$\tilde{f}_{ij}$ between them, to a limit space $S_{\infty}(F_{\infty})$ 
that 
is effectively homeomorphic to $S(X)$. The essential point to mention is 
that, since our 
constructive definition of the finitary sheaves $S_{n}$ in the 
previous section followed precisely, 
point-wise in $X$, the steps of the constructive definition of their 
respective domains $F_{n}$ in section 2, so that both $F_{n}$ and $S_{n}$ 
have the same 
poset-topology, one expects the proof of the convergence of ${\cal{N}}$ 
to $S(X)$ to be effectively the same as that of ${\cal{K}}$ to $X$ given 
in (Sorkin, 1991). Thus, our proof of the latter convergence only highlights 
the important definitions and proof-steps given by Sorkin (1991), referring 
the reader to it for more details.

The first thing from (Sorkin, 1991) that we mention is that `convergence' 
of ${\cal{K}}$ does not pertain to the usual notion of a `limit' for its 
terms, because for that, a topology on the set of all topologies on $X$ 
would have to exist, which is not the case. Rather, ${\cal{K}}$ is 
defined as an inverse system possessing an inverse limit.

The terms in ${\cal{K}}$ are the finitary substitutes of $X$ corresponding 
to the net ${\cal{L}}$ of locally finite open covers of $X$, together 
with unique continuous surjections 
$f_{ij}:~ F_{j}\rightarrow F_{i},~(i\preceq j)$\footnote{The uniqueness 
of the continuous $f_{ij}$s follows from the universal mapping theorem for 
$T_{0}$ topological spaces and the assumption that $i\preceq j$ 
(Sorkin, 1991).} between them. Thus, the inverse system ${\cal{N}}$ is 
defined in the same way as ${\cal{K}}$, though the continuous surjections between the 
finitary sheaves in it are now denoted by $\tilde{f}_{ij}$ ($i\preceq 
j$)\footnote{The `finer' relation $S_{i}\preceq S_{j}$ means in this case 
that the poset topology ${\cal{T}}(S_{i})$ is a subtopology of 
${\cal{T}}(S_{j})$, like for their corresponding finitary substitutes.}. 

As the inverse system ${\cal{K}}=<F_{j},f_{ij}>$ is seen 
to possess an inverse limit $(F_{\infty},f_{i\infty})$ 
as $j\rightarrow\infty$, with $F_{\infty}$ a 
$T_{0}$ topological space like all the finite terms in 
${\cal{K}}$\footnote{Again, the proof is again via the universal mapping 
property of the $f_{ij}$s.} (Sorkin, 1991), so does the inverse 
system ${\cal{N}}=<S_{j}(F_{j}),\tilde{f}_{ij}>$. The limit sheaf space 
$S_{\infty}(F_{\infty})$ is a $T_{0}$ topological space, since all the 
finite terms $S_{n}(F_{n})$ in ${\cal{N}}$ are $T_{0}$ posets.

The reader is referred to a series of lemmas in (Sorkin, 1991) that establish 
that the inverse limit space $F_{\infty}$ of the inverse system ${\cal{K}}$ 
is a non-Hausdorff space\footnote{A topological space $X$ is said to be Hausdorff, 
or satisfying the $T_{2}$ axiom of separation of point-set topology, if for 
every pair of distinct points 
$x$ and $y$ in it, there exist open 
neighborhoods $N(x)$ and $N(y)$ about them, such that $N(x)\cap 
N(y)=\emptyset$.} that contains $X$ as a dense subset. Since, as we 
noted earlier, the definitions and constructions for the terms in the 
inverse system ${\cal{N}}$ are identical, being implemented point-wise in 
$X$, with those in ${\cal{K}}$ of (Sorkin, 1991), we directly infer that 
the lemmas mentioned above also hold in our scheme, so that we may quote directly 
their result: the inverse limit $S_{\infty}(F_{\infty})$ of ${\cal{N}}$ 
is a non-Hausdorff space that contains $S(X)$ as a dense subset. The latter 
means essentially that for every stalk $S_{y}$ over $y\in F_{\infty}$ in the 
limit sheaf $S_{\infty}$ over $F_{\infty}$, there is a stalk $S_{x}$ in 
$f_{\infty}(S(X))$ `infinitely close' to it\footnote{One should read the 
lemmas in (Sorkin, 
1991) that establish this `infinite closeness' relation between points 
in $F_{\infty}$ and $f_{\infty}(X)$, 
and convince oneself that they also hold, point-wise, in our sheaf-theoretic 
scheme as well.}.  

Next, we mention a further lemma in (Sorkin, 1991) that establishes that if 
$X$ is $T_{1}$ and the previous lemmas for the denseness of $X$ in 
$F_{\infty}$ also hold, then $f_{\infty}(X)$ constitute the points of 
$F_{\infty}$ that are closed in its topology. We apply it to our situation and 
state that the image set $f_{\infty}(S(X))$ in $S_{\infty}(F_{\infty})$ 
consists of the closed stalks in the latter's topology. We refer again the reader to 
(Sorkin, 1991) to verify that $F_{\infty}$, and {\it in extenso} our 
$S_{\infty}(F_{\infty})$, is non-Hausdorff; moreover, one can `delete' its 
extra-points, thus render $f_{\infty}$ a homeomorphism between $S(X)$ and 
$S_{\infty}(F_{\infty})$. 

Finally, we mention the usefulness for the physical interpretation of 
the finitary spacetime sheaves $S_{n}(F_{n})$ as finite approximations 
of $S(X)$ in the next section, (the sheaf-theoretic analogue of Sorkin's 
proof) that the open subsets 
$\tilde{f}^{-1}_{i\infty}(S_{[x]})$ ($S_{[x]}\in S_{i}(F_{i})$) tiling  
$S_{\infty}(F_{\infty})$ become arbitrarily fine, or `small' as $j\rightarrow 
\infty$ (Sorkin, 1991)\footnote{See end of section 2.}. Thus, refining the finitary sheaves in ${\cal{N}}$ 
effectively amounts to better localizations or approximations of the observables 
residing in $S_{x}\in S(X)$\footnote{Read next section.}.} 

\section{\normalsize\bf Physical Significance and a Future Application of 
the ${\bf S_{n}}$s}

\sloppy{If we consider a net ${\cal{L}}=(\{{\cal{U}}_{n}\},\prec)$ of locally 
finite open covers of $X$, with ${\cal{U}}_{i}\prec{\cal{U}}_{j}$ denoting 
the relation `finer' 
between its elements as defined in section 2\footnote{Roughly, 
${\cal{U}}_{j}$ has more and 
`smaller' open subsets of $X$ than ${\cal{U}}_{i}$.}, then as shown 
in (Sorkin, 1991) and briefly discussed above, `in the limit of infinite 
resolution of $X$ into 
its `smallest' open neighborhoods (in ${\cal{U}}_{\infty}$) about its points, 
$X$ is recovered up to homeomorphism', that is to say, formally:  
$F_{\infty}({\cal{U}}_{\infty}):=\lim_{n\rightarrow\infty}F_{n}({\cal{U}}_{n})$ is homeomorphic to $X$.  
Thus, finitary substitutes are regarded as sound finite approximations of 
continuous topologies (Sorkin, 1991), whereby a `rough'\footnote{The 
epithets `fuzzy', `blurry', or `foamy' may be regarded as alternative 
synonyms to `rough'.}
determination of a point in $X$ is modelled after an open neighborhood about 
it (Sorkin, 1991, 1995). 

Now, the transcription of the problem `$F_{n}$ as approximations of $X$' to 
`$S_{n}(F_{n})$ as approximations of $S(X)$' that is the essence of the 
present paper, changes focus from `approximate 
localization/local determination (measurement) of points in 
$X$', to `approximate localization/local determination (measurement) of 
continuous functions over $X$', thus from a physical point of view, 
when $X$ is taken to be a bounded region of the spacetime manifold, 
from `localization of events', to `localization of observables of 
events'. In this paper, as it was mentioned earlier, the continuous topology 
of $X$ is regarded 
as its sole physically significant property `carried by its points',   
thus the continuous functions on it adequately qualify as `spacetime 
observables of events'. This attribute of points as `carrying the topology 
of $X$' can be realized by requiring that every physical space $X$ is a 
$T_{0}$ topological space (Sorkin, 1991). The relation 
$\stackrel{{\cal{U}}_{n}}{\sim}$ 
between events in $X$ is physically interpreted as `indistinguishability of 
events at the finite power of resolution of $X$ corresponding to 
${\cal{U}}_{n}$', while 
$\stackrel{{\cal{U}}_{n}}{\equiv}$ between the observables 
residing in the stalks of $S(X)$, as `indistinguishability of the observables 
of events at the finite power of resolution of $S(X)$ corresponding to 
${\cal{U}}_{n}$'. 

The posets corresponding to the locally finite substitutes $F_{n}$ of $X$ 
are known to have an equivalent ({\it ie}, functorial) representation as 
simplicial complexes obtained 
from the nerves of the covering ${\cal{U}}_{n}$ (Alexandrov, 1956, Raptis 
and Zapatrin, 2000) which, in 
turn, are categorically equivalent to incidence Rota 
algebras associated with them 
(Raptis and Zapatrin, 2000, Zapatrin, 2000)\footnote{That is to say, 
the category of finitary posets/order morphisms, is functorial to that of simplicial 
complexes/simplicial mappings, which, in turn, is `anti-functorial' to that of Rota 
algebras/Rota homomorphisms. The latter means that, since the Rota incidence 
algebras associated with finitary posets are objects dual to them (Raptis and 
Zapatrin, 2000), there is a contravariant functor between their respective 
categories. Thus, while finitary posets constitute the inverse system 
${\cal{K}}$ in (Sorkin, 1991) and above, their associated incidence algebras 
may be organized into a `direct system' having a direct limit, much like the 
finitary sheaves $S_{n}(F_{n})$ above were seen to have a direct limit space 
isomorphic to the stalk $S_{x}$ of $S(X)$.}. Observables too are in a certain sense dual 
to events\footnote{Intuitively they are dual, for the pairing of an 
observable $f$ with a point-event $x$ produces a measurable number 
$f(x)$({\it ie}, the value of the `field' $f$ at the `test-event' 
$x$). Thus, in some sense, $x$ is like a state of $X$, while the action of 
$f$ on it, $f(x)$, is some sort of measurement of the property $f$ of $X$ 
at $x$ (here, its continuous topology). Evidence for this duality is the 
mathematical duality of the notions of inverse and direct limit by which 
the local (point-like) elements of $X$ and $S(X)$ ({\it ie}, $x$ and 
$S_{x}$) were defined above, respectively. See also previous footnote.}. 
In (Raptis and Zapatrin, 2000) it was found that the Rota incidence algebra 
$R_{n}$ associated 
with a particular finitary substitute $F_{n}$ is also a discrete differential 
manifold in the sense of (Dimakis and M\"{u}ller-Hoissen, 1999). Thus,
not only the reticular analogues of the continuous ($C^{0}$) functions on 
$X$ are encoded in a finitary sheaf $R_{n}$ of Rota algebras over the $F_{n}$, but 
also a discrete version of the smooth ($C^{\infty}$) ones\footnote{Although,  
our study in the present paper concentrates solely on the continuous ({\it 
ie}, $C^{0}$) 
structure of spacetime, not its differential.}. Hence, the conjecture 
is that at the maximum resolution ${\cal{U}}_{\infty}$ of a bounded 
region $X$ of a smooth spacetime manifold $M$, the Rota 
algebras $R_{n}$ associated with the locally finite posets $F_{n}$, 
$R_{n}(F_{n})$, are expected to `yield' $(X,\partial 
,\omg)$\footnote{$\omg$ being the module of differential 
forms over the algebra of $C^{\infty}$-smooth functions on $X$, {\it ie}, 
$\omg:=\omg^{0}(\equiv C^{\infty}(X)) \oplus \omg^1 \oplus \omg^{2}\oplus
\ldots$}-the flat 
sheaf of sections of smooth differential forms over $X$ ({\it ie}, the 
smooth spacetime observables)\footnote{In (Mallios, 1998), this structure 
is called the `smooth and flat 
differential triad'. The stalks of this sheaf are isomorphs of $\omg$ and 
the K\"ahler-Cartan differential $\partial$ effects (stalk-wise) vector 
sheaf morphisms of the following sort: 
$\partial:~(X,\omg^{n})\rightarrow(X,\omg^{n+1})$; where $(X,\omg^{n})$ is 
the vector subsheaf of $(X,\partial ,\omg)$ having as stalks isomorphs of 
the vector space 
$\omg^{n}$ of $n$-forms, which is a vector subspace of $\omg$.}. Also, 
a quantum interpretation has been given to the Rota incidence 
algebras $R_{n}$ associated with the finitary substitutes $F_{n}$ of $X$ 
(Raptis and Zapatrin, 2000). Accordingly, the limit space of a system of 
$R_{n}(F_{n})$s is $(X,\partial ,\omg)$-the classical smooth spacetime 
manifold with differential forms attached, and it was 
interpreted there as Bohr's correspondence limit structure of a quantal 
substratum of finitary incidence algebras.

In (Raptis, 2000), a causal interpretation to the quantal incidence 
algebras $R_{n}(F_{n})$ of (Raptis and Zapatrin, 2000) was given. The resulting 
structures were called `quantum causal sets'-a quantal version of the causal 
sets of (Bombelli {\it et al.}, 1987). It follows from the discussion above that a finitary  
sheaf of quantum causal sets may be studied as a quantal and locally 
finite substitute of the causal relations between events in a bounded region of a 
smooth Lorentzian spacetime manifold. As in 
(Dimakis and M\"{u}ller-Hoissen, 1999) a Riemannian metric connection was studied on a 
discrete differential manifold, so we should be able to define a 
non-flat pseudo-Riemannian connection on the finitary sheaf of quantum 
causal sets by using powerful sheaf-theoretic results from (Mallios, 
1998)\footnote{These results were obtained for a 
paracompact and Hausdorff topological base space $X$ of the vector sheaves 
considered there. A space that is paracompact is akin 
to one that is finitary in our and Sorkin's (1991) sense of the latter 
denomination ({\it ie}, that it admits a locally finite open cover). Our $X$, which 
is such a finitary topological space, was  
assumed to be bounded ({\it ie}, having compact closure), and in the usual 
topological parlance it is called `relatively compact'. $X$ was also seen to 
be $T_{1}$, but not $T_{2}$ ({\it ie}, not Hausdorff). 
If we relax `paracompactness' to `relative compactness' and $T_{2}$ to 
$T_{1}$ for $X$, the essential results of Abstract Differential Geometry via 
Vector Sheaves still apply to our case (Mallios, private communication).}.  
It is expected that in the Bohr correspondence limit\footnote{In the sense of 
(Raptis and Zapatrin, 2000).} of an inverse system of finitary 
sheaves of quantum causal sets, the smooth (region of a) Lorentzian manifold, 
together with the smooth fields and a non-flat pseudo-Riemannian metric 
connection $D$ on it\footnote{Interpreted as the classical gravitational potential.}, 
will emerge. Then, the underlying finitary sheaves of 
quantum causal sets may be regarded as sound models of a reticular and quantal 
version of gravity\footnote{A `finitary quantum gravity' so to speak.}. This 
project however, is still under development (Mallios and Raptis, 2000). 

We conclude this paper by discussing its `general physico-philosophical 
moral'. By considering the finitary sheaves approximating $S(X)$ rather than 
directly the finitary spaces approximating $X$, we regard our observations 
of spacetime events as being more fundamental than the events themselves. 
This is the main lesson for physics to learn by applying 
differential geometry in the inherently 
algebraic language of sheaf theory, namely, that the 
physically significant concepts are less those about the `geometrical' background 
spacetime $X$ and more those about our observations of this background which 
are organized into sheaves (of algebras) over $X$ (Mallios, 1998). 
This general principle that underlies the Abstract Differential Geometry via 
Vector Sheaves theory developed in (Mallios, 1998), is well in accord with 
the general philosophy of quantum theory holding that inert, background, 
geometrical 
`state spaces', such as spacetime, `dissolve away', so that what remains and is 
of physical significance, the `physically real' so to speak, is (the 
algebraic mechanism of) our own 
actions of observing `it' (Finkelstein, 1996)\footnote{To parallel Saunders 
Mac Lane's 
mathematical motto `Every good function is a section of a sheaf' (Mac Lane, 
1986) in physics: `Every physically significant action is a section of the 
sheaf of our operations of observation of the system in focus'.}.}

\section*{\normalsize\bf Acknowledgments}

\sloppy{The incessant support: spiritual, moral, technical and material, of Anastasios 
Mallios, is wholeheartedly acknowledged. Exchanges on finitary spacetime 
substitutes during a year of collaboration with Roman Zapatrin, are also 
greatly appreciated. Finally, this paper was written with the help of a 
post-doctoral research fellowship in mathematics from the University of 
Pretoria.}       

\section*{\normalsize\bf References}

\sloppy{\noindent Alexandrov, P. S. (1956). {\it Combinatorial Topology}, vol. 1,  
Greylock, Rochester, New York.

\noindent Bredon, G. E. (1967). {\it Sheaf Theory}, McGraw-Hill, New York.

\noindent Bombelli, L., Lee, J., Meyer, D. and Sorkin, R. D. (1987). 
{\it Physical Review Letters}, {\bf 59}, 521.

\noindent Dimakis, A. and M\"uller-Hoissen, F. (1999). {\it Journal of 
Mathematical Physics}, {\bf 40}, 1518.

\noindent Finkelstein, D. (1996). {\it Quantum Relativity}, Springer Verlag, 
Berlin-Heidelberg-New York.

\noindent Mac Lane, S. (1986). {\it Mathematics: Form and Function}, Springer 
Verlag, New York.

\noindent Mallios, A. (1998). {\it Geometry of Vector Sheaves}, 
vols. 1-2, Kluwer Academic Publishers, Dordrecht.
                               
\noindent Mallios, A. and Raptis, I. (2000). {\it Finitary Spacetime Sheaves of Quantum Causal 
Sets: Curving Quantum Causality}, in preparation.

\noindent Raptis, I. (2000). {\it International Journal of Theoretical Physics}, {\bf 39}, 1233.

\noindent Raptis, I. and Zapatrin, R. R. (2000). {\it International Journal of Theoretical
Physics}, {\bf 39}, 1.

\noindent Sorkin, R. D. (1991). {\it International Journal of Theoretical 
Physics}, {\bf 30}, 923.

\noindent Sorkin, R. D. (1995). {\it A Specimen of Theory Construction from 
Quantum Gravity} in {\it The Creation 
of Ideas in Physics}, ed. Leplin, J., Kluwer Academic Publishers, Dordrecht.

\noindent Zapatrin, R. R. (2000). {\it Incidence algebras of simplicial 
complexes}, paper submitted to {\it Pure Mathematics and its Applications}, e-print 
math.CO/0001065.}

\end{document}